\documentclass[twocolumn,prb,floatfix,amssymb,aps, showpacs]{revtex4}
\usepackage{graphicx}
\usepackage{color}
\usepackage{epsfig}
\usepackage{amsmath}
\usepackage{amssymb}
\usepackage{mathrsfs}
\usepackage{subfigure}
\usepackage{url}
\usepackage{enumerate}
\usepackage{braket}
\usepackage{comment}

\begin{document}

\hsize\textwidth\columnwidth\hsize\csname@twocolumnfalse\endcsname

\title{The evolution of magnetic structure driven by a synthetic spin-orbit coupling in two-component Bose-Hubbard model}
\author{Jize Zhao$^{1,2}$}\email{jizezhao@gmail.com}
\author{Shijie Hu$^3$, Jun Chang$^4$, Fawei Zheng$^{1,2}$, Ping Zhang$^{1,2}$, Xiaoqun Wang$^{5,6}$}
\affiliation{$^1$Institute of Applied Physics and Computational Mathematics, Beijing 100088, China}
\affiliation{$^2$Beijing Computational Science Research Center, Beijing 100084, China}
\affiliation{$^3$Max-Plank Institute f\"ur Physik Komplexer Systeme, Dresden 01187,  Germany}
\affiliation{$^4$College of Physics and Information Technology, Shaanxi Normal University, Xi'an 710062, China}
\affiliation{$^5$Department of Physics and Astronomy, Shanghai Jiao Tong University, Shanghai 200240, China}
\affiliation{$^6$Department of Physics and Beijing Laboratory of of Opto-electronic Functional Materials$\&$Micro-nano Devices, Renmin University of China, Beijing, 100872, China}
\begin{abstract}
We study the evolution of magnetic structure driven by a synthetic spin-orbit coupling 
in a one-dimensional two-component Bose-Hubbard model. In addition to the Mott insulator-superfluid transition, 
we found in Mott insulator phases a transition from a gapped 
ferromagnetic phase to a gapless chiral phase by increasing the strength of spin-orbit coupling. 
Further increasing the spin-orbit coupling drives a transition from the gapless chiral phase to 
a gapped antiferromagnetic phase. These magnetic structures persist in superfluid phases. 
In particular, in the chiral Mott insulator and chiral superfluid phases, incommensurability  
is observed in characteristic correlation functions.  
These unconventional Mott insulator phase and superfluid phase demonstrate the novel effects arising from the 
competition between the kinetic energy and the spin-orbit coupling. 
\end{abstract}

\maketitle

\section{INTRODUCTION}
Spin-orbit coupling (SOC) is known to play an important role for many exotic phenomena in condensed matter physics,
such as topological insulators\cite{KANE1, QI1, HASAN1, QI2}, topological superconductivity\cite{QI2,FU1} and 
unconventional magnetism\cite{BANERJEE1}. Example materials include HgCdTe quantum well\cite{BERNEVIG1}, 
Copper Benzoate\cite{BROHOLM1, OSHIKAWA1, ZHAO1}. In solid materials, SOC originates from 
relativistic correction and widely exists in crystals with low symmetry. However, it is
very weak in comparison with dominant energy scales and usually considered as a perturbation\cite{DZ,MORIYA}. Moreover, its form 
strongly depends on the internal structure of the materials and is thus difficult to be tuned in experiments.  
These facts hinder our further understanding of the SOC and related phenomena so far. 

Recently, the realization of a synthetic SOC in ultracold atomic systems made a substantial progress
towards overcoming these difficulties. With a pair of tunable lasers by dressing 
two internal atomic states\cite{LIN1}, Lin {\it{et al}}. engineered a SOC with equal Rashba and Dresselhaus weight 
in neutral atomic Bose-Einstein condensate. Via modifying the interaction between two atomic states, 
a quantum phase transition is observed. Although the effects of the SOC were found long time ago in solid materials, 
this is the first experimental observation of the SOC in bosonic systems and its properties have not been fully explored yet.  
Soon after that, the same type of the SOC in fermion systems was realized in ultracold atomic experiments\cite{WANG1, CHEUK1} by 
the similar technique. The advantages of this synthetic SOC is that its strength can be 
adjusted in experiments\cite{GALITSKI1}, allowing full access to the parameter region we are interested in. 
These experiments pave the way for deepening our understanding of the SOC in both ultracold 
atomic systems and condensed matter physics. 

Up to now, whereas most experimental and theoretical studies on the SOC mainly focus on 
continuous systems\cite{LIN1, WANG1, CHEUK1, GALITSKI1, WU1, ZHOU1, HU1, YANG1, ISKIN1, DONG1, ZHAI1, ZHOU2},
some experimentalists and theorists recently turn to study related phenomena in optical lattices. 
On one hand, lattices are the basic structure in solid state physics and  
optical lattices are the key for simulating condensed matter physics\cite{DUAN1, FUKUHARA1} in ultracold atomic setups. 
On the other hand, the advantage of lattice systems over continuum is that a variety of  analytical 
and numerical methods for studying strongly correlated systems have been well established in the past several decades.  
For examples, theoretical works based on spin wave, slave boson and Monte carlo have predicted a rich zero-temperature 
magnetic phase diagram\cite{MANDAL1, CAI1, GONG1, RADIC1, COLE1} 
for two-dimensional bosonic systems with Rashba SOC in the Mott region. It includes 
ferromagnetic (FM), antiferromagnetic (AF), spiral, vortex and Skyrmion phases.
More recently, unconventional superfluid (SF) phases have been found\cite{QIAN1, BOLUKBASI1} in the Bose-Hubbard model 
with various types of SOC in two dimensions. These results are in contrast to the relatively 
simple Mott insulator (MI)  and SF phases for the two-component Bose-Hubbard model\cite{ALTMAN1} without the SOC. 
In addition, fermionic lattice models with the Rashba SOC recently also become the context of theoretical investigations 
of some intriguing effects in various geometries such as ladders\cite{RIERA1} and square lattice\cite{TANG1}.  

In spite of those theoretical great interests in the Rashba SOC recently\cite{GALITSKI1,XU1}, there is no experimental 
realization of such a SOC in ultracold atomic systems so far. Therefore, in this paper we focus on the SOC that 
has already been realized in ultracold atom experiments, 
which is proportional to $p_x\sigma_y$. Since this SOC is along one spatial direction and of abelian nature,
it is natural to start our study on the effects of the SOC from a one-dimensional model.
In the tight-binding form, Hamiltonian including the kinetic energy and the SOC can be generally written as 
\begin{eqnarray}
\mathcal{H}_{t\lambda} = -t\sum_{i\tau }(\hat{c}_{i\tau}^{\dagger}\hat{c}_{i+1\tau}+h.c.)+\mathcal{T}_{soc}  
\label{EHTL}
\end{eqnarray}
where $\hat{c}_{i\tau} (\hat{c}^{\dagger}_{i\tau})$ denotes the annihilation (creation) operators
at site $i$ for spin $\tau$. $\tau$ takes $\uparrow$ and
$\downarrow$, representing two internal states of atoms. The first term in Eq. (\ref{EHTL}) is the kinetic energy and 
$t$ is the hopping matrix between nearest neighbor sites. 
$\mathcal{T}_{soc}$ describes the SOC realized in experiments. Its tight-binding form is represented by 
\begin{equation}
\mathcal{T}_{soc} = -\lambda\sum_i(\hat{c}^{\dagger}_{i\uparrow}\hat{c}_{i+1\downarrow} -\hat{c}^{\dagger}_{i\downarrow}\hat{c}_{i+1\uparrow})+h.c.,
\end{equation}
with $\lambda$ being a SOC strength.

It turns out that $\mathcal{T}_{soc}$ can be eliminated by a site-dependent rotation in its internal space of each site, 
resulting in a renormalization of the hopping integral $t$. 
To show this, we take the following rotation\cite{SHEKHTMAN1,CAI1} around y-axis at site $i$,
\begin{eqnarray}
\left(\begin{array}{c}
\hat{c}_{i\uparrow}\\
\hat{c}_{i\downarrow}
\end{array}\right) & = & \left(\begin{array}{cc}
\cos\frac{\omega_i}{2} & -\sin\frac{\omega_i}{2}\\
\sin\frac{\omega_i}{2} & \cos\frac{\omega_i}{2}
\end{array}\right)\left(\begin{array}{c}
\hat{c}_{i\uparrow}^{\prime}\\
\hat{c}_{i\downarrow}^{\prime}
\end{array}\right).
\label{EROTATE}
\end{eqnarray}
By substituting Eq. (\ref{EROTATE}) into Eq. (\ref{EHTL}), and setting $\omega_{i+1}-\omega_i = 2\arctan(\frac{\lambda}{t})$,
one finds that $\mathcal{T}_{soc}$ disappears with $t$ renormalized to $\sqrt{t^2+\lambda^2}$. 
However, a realistic Hamiltonian inevitably involves some other terms 
such as interactions for correlations or a Zeeman term for experiments, which we denote as $\mathcal{H}_{other}$. 
If $\mathcal{H}_{other}$ is invariant under the rotation (\ref{EROTATE}), the net effect of SOC is merely 
renormalizing the hopping term and considered to be trivial in this case. The situation can be easily altered by 
introducing an (artificial) Zeeman field into $\mathcal{H}_{other}$. In this case, 
the SOC becomes relevant and new physical phenomena
are expected to occur. In particular, in the presence of both the SOC and Zeeman field, 
a variety of exotic phases such as topological superconductivity
have been proposed\cite{QU1,ZHANG1,LIU1, CAO1} in fermion systems. 

Zeeman field definitely breaks the time-reversal symmetry and  
the polarization strongly depends on the direction of the Zeeman field. 
Instead, in our work, we consider an experimentally relevant model 
without breaking time-reversal symmetry, i.e. a two-component Bose-Hubbard model with SOC, which reads 
\begin{eqnarray}
\mathcal{H} & = & \mathcal{H}_{t\lambda}+\frac{U}{2}\sum_{i\tau} \hat{n}_{i\tau}(\hat{n}_{i\tau}-1)
                  + U^{\prime} \sum_i \hat{n}_{i\uparrow} \hat{n}_{i\downarrow} \nonumber \\
            &   & -\mu\sum_{i}(\hat n_{i\uparrow}+\hat n_{i\downarrow}),
\label{HSOC}
\end{eqnarray}
where $\mathcal{H}_{t\lambda}$ is the same as Hamiltonian (\ref{EHTL}) but $\hat{c}_{i\tau}$, $\hat{c}^{\dagger}_{i\tau}$ 
is restricted for bosons here. 
$U$ is on-site intracomponent interaction and $U^\prime$ is the intercomponent one. 
$\hat{n}_{i\tau}=\hat{c}^\dagger_{i\tau}\hat{c}_{i\tau}$ is the boson number operator with spin $\tau$ at site $i$.
$\mu$ is the chemical potential to control the filling factor. When $U^\prime=U$, the interaction part is invariant 
under the rotation (\ref{EROTATE}). Therefore, $\mathcal{T}_{soc}$ can be eliminated\cite{CAI1}, resulting in a standard 
two-component Bose-Hubbard model (TBHM), which has been extensively studied in 
literature\cite{ALTMAN1, MISHRA1, HU2, ISACSSON1}. However, when $U^\prime \ne U$,
the interaction part is not invariant any more under the rotation (\ref{EROTATE}). 
One can then expect that some new phenomena would emerge. For example, spontaneous $Z_2$ symmetry breaking resulting from   
the competition between the kinetic energy and SOC has been predicted\cite{ZHAO2} for $U^\prime<U$.   
 
We recall that Hamiltonian (\ref{HSOC}) has a $U(1)\times Z_2$ symmetry\cite{ZHAO2}, described by the transformation
\begin{eqnarray}
\hat c_{i\tau^{\prime}} \rightarrow \sum_\tau [e^{i\phi} e^{-i\pi\sigma_y/2} ]_{\tau^{\prime} \tau} \hat c_{i\tau},
\end{eqnarray}
which indicates that only the total particle number is conserved. Moreover, Hamiltonian (\ref{HSOC}) is unchanged by 
interchanging $t$ and $\lambda$ with the following transformation
\begin{eqnarray}
&&\hat c_{i \tau}~~~\rightarrow ~~{\rm sign}_\tau ~\hat c_{i \tau}, \ \  \ \ ~\hat c_{i+1 \tau} \rightarrow ~~\hat c_{i+1 \bar{\tau}}, \nonumber  \\
&&\hat c_{i+2 \tau} \rightarrow -{\rm sign}_\tau ~\hat c_{i+2 \tau}, \ \ \hat c_{i+3 \tau} \rightarrow -\hat c_{i+3 \bar{\tau}} 
\label{EXCHANGE}
\end{eqnarray}
for every 4-sites with ${\rm sign}_\uparrow=1$ and ${\rm sign}_\downarrow=-1$ and $\bar \tau$ represents the
opposite spin of $\tau$. These symmetries would be helpful for understanding the properties of Hamiltonian (\ref{HSOC}).
Particularly, we can define the ratio 
\begin{equation}
\eta=\lambda/(t+\lambda)
\label{ETA}
\end{equation} 
to establish a symmetric phase diagram with respect to the axis $\eta=0.5$. The phases with $\eta\in(0.5,1]$ can be readily 
figured out from those with $\eta\in[0,0.5]$ in terms of the transformation (\ref{EXCHANGE}).

In the absence of $\mathcal{T}_{soc}$, Hamiltonian (\ref{HSOC}) is reduced to the well-known TBHM. 
In large $U, U^\prime$ limit and at a unit filling, it can 
be mapped to a spin-1/2 XXZ Heisenberg model. In particular, when $U^\prime > U$, its ground state is 
in a z-axis Ising FM\cite{ALTMAN1} phase.
Consequently, its low-energy excitation is gapful. When a small $\mathcal{T}_{soc}$ is set in, 
we would expect that its ground state remains in such a gapped FM phase.
This thus greatly motivates us in this work to explore how this magnetic order evolves as the SOC increases continuously.

The rest of the paper is organized as follows: In Section II, we derive an effective 
magnetic Hamiltonian in large $U,U^\prime$ limit at a unit filling. 
We first discuss some special cases, where the effective Hamiltonian can be simplified further
so that we are able to give qualitative conclusions with available analytic results.  
Then we present an accurate phase diagram which is established numerically by means of 
the density-matrix renormalization group (DMRG)\cite{WHITE1,PESCHEL1,SCHOLLWOCK1} method.
In Section III, we study the transition between the MI and the SF phases, 
the magnetic structures as well as the momentum distribution in SF phases. 
Particularly, we focus on the evolution of the magnetic structure with respect to the SOC in SF phases.   
In Section IV, we give our conclusions.
 
\section{EFFECTIVE MAGNETIC HAMILTONIAN IN DEEP MOTT REGION}
In this section, we discuss the magnetic properties in the MI phases at the unit filling.
To understand the magnetic properties of Hamiltonian (\ref{HSOC}), it is natural to start our discussion from the deep MI 
region, i.e., $t, \lambda \ll U, U^{\prime}$, where charge freedom is frozen and only spin freedom is involved at zero temperature.  
Under such a restriction, the low-energy behavior of Hamiltonian (\ref{HSOC}) can be effectively described by a spin-1/2 model. 
To see this, we split the Hamiltonian into two parts, $\mathcal{H} = \mathcal{H}_{t\lambda}+\mathcal{H}_{other}$, where 
\begin{eqnarray} 
\mathcal{H}_{other} = \frac{U}{2}\sum_{i\tau}\hat{n}_{i\tau}\left(\hat{n}_{i\tau}-1\right)+U^{'}\sum_{i}\hat{n}_{i\uparrow}\hat{n}_{i\downarrow}.
\end{eqnarray}
Here, $\mathcal{H}_{t\lambda}$ includes both the kinetic energy and the SOC and is taken as a perturbation. 
The ground states of $\mathcal{H}_{other}$ have one particle per site and is thus highly degenerate. To the second order, 
the effective Hamiltonian can be derived as 
\begin{widetext}
\begin{multline}
\mathcal{H}_{eff}  =  \mathcal{P}\mathcal{H}_{t\lambda} \frac{1}{E_0-\mathcal{H}_{other}}(1-\mathcal{P})\mathcal{H}_{t\lambda}\mathcal{P} 
                   =  \left(\frac{8\left(t^{2}-\lambda^{2}\right)}{-U}+\frac{4\left(\lambda^{2}-t^{2}\right)}{-U^{'}}\right)\sum_{i}\hat{S}_{i}^{z}\hat{S}_{i+1}^{z}+\frac{4\left(t^{2}-\lambda^{2}\right)}{-U^{'}}\sum_{i}\hat{S}_{i}^{x}\hat{S}_{i+1}^{x}\\
 +\frac{4\left(t^{2}+\lambda^{2}\right)}{-U^{'}}\sum_{i}\hat{S}_{i}^{y}\hat{S}_{i+1}^{y}
-\frac{8\lambda t}{U}\sum_{i}\left(\hat{S}_{i}^{z}\hat{S}_{i+1}^{x}-\hat{S}_{i}^{x}\hat{S}_{i+1}^{z}\right)
\label{EHEFF}
\end{multline}
\end{widetext}
where $\mathcal{P}$ is the projection operator onto the subspace spanned by the ground states of $\mathcal{H}_{other}$, 
$E_0$ is the ground state energy of $\mathcal{H}_{other}$, and  
\begin{eqnarray}
\hat{S}_{i}^{\nu}= \sum_{\tau\tau^\prime}\hat{c}_{i\tau}^\dagger\sigma^{\nu}_{\tau\tau^\prime}\hat{c}_{i\tau^\prime}/2
\label{SPIN}
\end{eqnarray}
are the pseudo-spin operators with $\nu=x,y,z$ and $\sigma^{\nu}$ Pauli matrix. To derive Hamiltonian (\ref{EHEFF}), we use the relations
$(\hat{n}_{i\uparrow}+\hat{n}_{i\downarrow})(\hat{n}_{i+1\uparrow}+\hat{n}_{i+1\downarrow})=1$, 
and $(\hat{n}_{i\uparrow}-\hat{n}_{i\downarrow})(\hat{n}_{i+1\uparrow}-\hat{n}_{i+1\downarrow})=4\hat{S}_i^z\hat{S}_{i+1}^z$.
The Hamiltonian (\ref{EHEFF}) is just a spin-1/2 XYZ Heisenberg model with the Dzyaloshinskii-Moriya interaction\cite{DZ,MORIYA}. 
In some special cases, this Hamiltonian can be mapped\cite{OSHIKAWA1} into an exactly solvable model.  
This would definitely shed light on our understanding of the properties of Hamiltonian (\ref{HSOC}).
First, when $\lambda=0$, the ground state of Eq. (\ref{EHEFF}) is a gapped Ising FM state 
with the polarization in $z$ direction. Second, when $t=0$, Eq. (\ref{EHEFF}) has a gapped AF ground state, 
which is readily known through the transformation (\ref{EXCHANGE}). 
Thirdly, when $t=\lambda$, it is in a critical phase with gapless excitations. 
An important implication from above analysis is that there are at least two transition points 
when $\eta$ varies from 0 to 1.  

\begin{figure}
\includegraphics[width=8.8cm, clip]{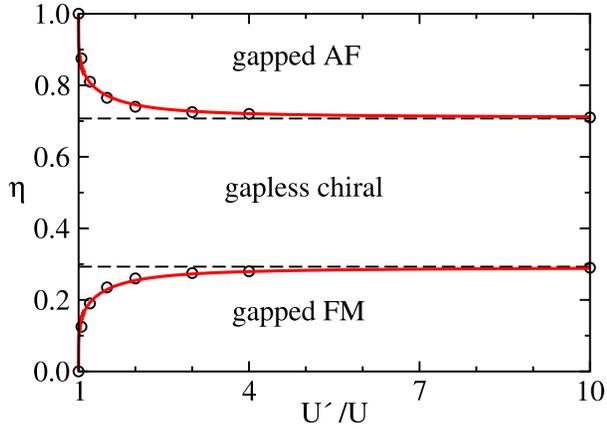}
\caption{(Color online) Magnetic phase diagram is shown in a deep Mott region. 
The open circles are the phase boundary obtained by DMRG.
The lower dashed line is $\eta=\eta_{\scriptscriptstyle{\infty}}$ and 
the upper dashed line is $\eta=1-\eta_{\scriptscriptstyle{\infty}}$ with $\eta_{\scriptscriptstyle{\infty}}=1-\sqrt{2}/2$,
corresponding to the two critical points at $U^\prime/U=\infty$.
The red solid lines are fitting for the numerical data, given by the 
function $\eta=\eta_{\scriptscriptstyle{\infty}}\left(\frac{2}{\pi}\arctan(\frac{U^\prime}{U}-1)\right)^{1/5}$ for the lower boundary and
$\eta=1-\eta_{\scriptscriptstyle{\infty}}\left(\frac{2}{\pi}\arctan(\frac{U^\prime}{U}-1)\right)^{1/5}$ for the upper boundary.}
\label{FIG1}
\end{figure}
To establish an accurate magnetic phase diagram, we employ the DMRG method to study the Hamiltonian (\ref{HSOC}).
In our calculation, we take $U$ as the energy unit and fix $t+\lambda=0.04U$,
which well satisfy the above constraint $t,\lambda\ll{U,U^\prime}$.
In Fig. \ref{FIG1}, we show the magnetic phase diagram in $U^\prime/U-\eta$ plane in a deep Mott insulator region.
One can see that the phase diagram consists of a gapped FM phase, a gapless chiral phase and 
a gapped AF phase from the bottom to the top. In both the gapped FM and the gapped AF phases, the polarization is in z-axis.
The phase boundary has a reflection symmetry with respect to the axis $\eta=0.5$ owing to the transformation (\ref{EXCHANGE}).
According to our previous analysis based on the effective Hamiltonian (\ref{EHEFF}), we know that $\eta=0$ corresponds to
a gapped FM phase, which is in agreement with our numerical results. 
Such a gapped FM phase extends to a finite $\eta_c$, with $\eta_c$ depending monotonically on 
$U^\prime/U$. Further increasing $\eta$ drives the ground state into a gapless chiral phase
and then transits from the gapless chiral phase into a gapped AF phase at $1-\eta_c$ and stays in the gapped AF phase for $\eta$ up to 1.
When $U^\prime/U=\infty$, $\eta_c$ can be obtained\cite{DERZHKO1} exactly as $\eta_c=\eta_{\infty}=1-\sqrt{2}/2$. 
Furthermore, we found that the two phase boundaries can be well fitted by a elegant formula 
$\eta=\eta_{\scriptscriptstyle{\infty}}\left(\frac{2}{\pi}\arctan(\frac{U^\prime}{U}-1)\right)^{\alpha}$ 
and $\eta=1-\eta_{\scriptscriptstyle{\infty}}\left(\frac{2}{\pi}\arctan(\frac{U^\prime}{U}-1)\right)^{\alpha}$, 
respectively, with $\alpha=1/5$. It is worthwhile to call for analytic works, for example, by bosonization technique,
to gain further insightful information related to these interesting results.

Now let us discuss the difference among these three phases in terms of spin-spin correlation functions, chiral 
correlation functions and excitation gaps. 
For simplicity, we restrict our discussion to the case $U^\prime/U=1.2$, since the results at other $U^\prime/U$ 
are found qualitatively the same.  
First, we found that the three phases are clearly distinguished
by the spin-spin correlation functions defined as
\begin{equation}
\mathcal{S}^{\nu}_{ij} = \langle\psi_0|\hat S^{\nu}_{i} \hat S^{\nu}_{j}|\psi_0\rangle, 
\label{SCORR}
\end{equation}
and the corresponding static structure factors
\begin{equation}
\mathcal{S}^\nu\left(q\right)=\frac{1}{L}\sum_{ij}e^{iq\left(i-j\right)}S^\nu_{ij},
\label{SQCORR}
\end{equation}
where $q$ is the momentum, $|\psi_0\rangle$ is the ground state of Hamiltonian (\ref{HSOC}) and $L$ is the chain length.
\begin{figure}
\includegraphics[width=8.6cm, clip]{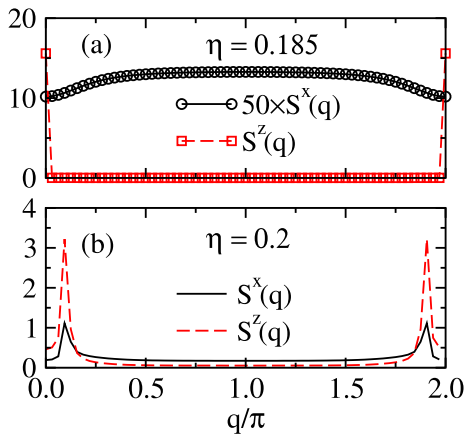}
\caption{(Color online) Typical static structure factors as a function of $q$ are shown (a) in the gapped FM phase at $\eta=0.185$ 
and (b) in the gapless chiral phase at $\eta=0.2$. In panel (a), $S^{x}(q)$ is scaled by 50 for clear vision. 
The chain length $L$ is 64.}
\label{FIG2}
\end{figure}
Hereafter, our discussion on $q$-dependence of the structure factors are restricted to $q\in[0,\pi]$ 
since $S^\nu(q)$ is symmetric with respect to $q=\pi$ (and the restriction is also applied to the momentum distribution in the next section).
In Fig. \ref{FIG2} (a), the static structure factor $S^{x}(q)$ and $S^{z}(q)$ are shown for the gapped FM phase. 
We found that a peak shows up at $q=0$ in $S^z(q)$. Moreover, $S^z(0)/L$ scales to a finite value 
in the thermodynamic limit, indicating a long-range FM order. However, it is quite different in 
the gapless chiral phase. As we show in Fig. \ref{FIG2} (b), 
the peaks of both $S^x(q)$ and $S^z(q)$ are shown at the same $q$ but with $q\ne{0,\pi}$, reflecting an incommensurate structure factor. 
This incommensurability uncovers a characteristic feature of the gapless chiral phase.
Moreover, Fig. \ref{FIG2} indicates that a phase transition occurs between $\eta=0.185$ and $\eta=0.2$.
\begin{figure}
\includegraphics[width=9cm,clip]{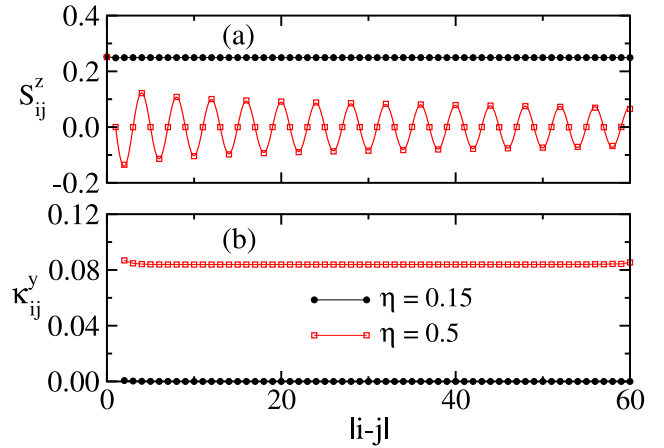}
\caption{(Color online) Typical correlation functions $S^z_{ij}$ and $\kappa_{ij}^y$ at $t+\lambda=0.04U$
as a function of $|i-j|$ are shown in the gapped FM phase ($\eta = 0.15$)
and gapless chiral phase ($\eta = 0.5$) for $L=128$. We fix $i$ at the middle site of the chain. 
Four sites near the edge are not shown to avoid edge effect.
The red solid line in panel (a) is fitting for those data obtained by DMRG at $\eta=0.5$
with $S^z_{ij}=0.1614\times\cos(|i-j|\pi/2)/|i-j|^{0.197}$. 
This figure clearly demonstrates the existence of long-range chiral order in the gapless chiral phase and FM order
in the gapped FM phase.}
\label{FIG3}
\end{figure}
In order to further illustrate different features between these two phases, 
we plot in Fig. \ref{FIG3} (a) $S^z_{ij}$ as a function of $|i-j|$ at 
$\eta=0.15$ for the gapped FM phase and at $\eta=0.5$ for the gapless chiral phase. 
These two values of $\eta$ are chosen a little far from the critical point 
to gain a clear discrimination between the two phases for the typical size we studied. As expected,
$S^z_{ij}$ is finite in the large $|i-j|$ limit in the gapped FM phase, while it decays algebraically 
in the gapless chiral phase as a function of $|i-j|$, which can be well fitted by
\begin{eqnarray}
S^z_{ij}\sim\cos(|i-j|q_{s}+\delta_s)/|i-j|^{\alpha_s},
\label{SZZFIT}
\end{eqnarray}
where $q_s$, $\delta_s$ and $\alpha_s$ depend on $(t+\lambda)/U$, $\eta$, and $U^\prime/U$. Moreover, $q_s$ just corresponds to the
momentum where the peak of $S^z(q)$ locates, and depends monotonically on $\eta$. In particular, at $\eta=0.5$, $q_s=\pi/2$, 
which is independent of other parameters but as an immediate consequence of the symmetry revealed by the transformation (\ref{EXCHANGE}). 
This power-law behavior of $S^z_{ij}$ is reminiscent of Tomonaga-Luttinger liquid in the gapless chiral phase.  

Second, the difference between the gapless chiral phase and other phases can be detected by chiral
correlation functions, defined by
\begin{equation}
\mathcal{\kappa}^\nu_{ij} = \langle\psi_0|({\hat{\bf{S}}_i}\times{\hat{\bf{S}}_{i+1}})^\nu({\hat{\bf{S}}_j}\times{\hat{\bf{S}}_{j+1}})^\nu|\psi_0\rangle,\ \nu=x,y,z. 
\label{CHIRAL}
\end{equation}
It demonstrates completely different behavior for the gapless chiral phase from the gapped FM phase as shown in Fig. \ref{FIG3} (b).
We plot $\mathcal{\kappa}^y_{ij}$ as a function of $|i-j|$ at $\eta = 0.15$
(for the gapped FM phase) and at $\eta = 0.5$ (for the gapless chiral phase).
$\kappa^y_{ij}$ is finite in the large $|i-j|$ limit for the gapless chiral phase, indicating the existence of long-range chiral order.
However, for the gapped FM phase, it is exponentially small.
We also checked other $\eta$ and confirmed that in the gapped FM phase $\mathcal{\kappa}^y_{ij}$ is always exponentially small,
and in the gapless chiral phase it is always finite in the large $|i-j|$ limit.
However, near the transition point but in the gapless chiral phase, oscillation around a finite value is observed.
We also calculated $\kappa^x_{ij}$, $\kappa^z_{ij}$ and found that they are always exponentially 
small so we will not discuss them further. Our results clearly demonstrate the long-range chiral order in the gapless chiral phase.

Finally, these phases can be distinguished by a longitudinal gap, defined by
\begin{equation}
\Delta_k=E_k(N,L)-E_0(N,L),
\end{equation}
where $E_k(N,L)$ is the energy of $k$-th excited state with particle number $N$ and length $L$ and 
$E_0(N,L)$ is the energy of the ground state with particle number $N$ and length $L$.
In our work, we have to $k=2$ due to the two-fold degeneracy of the ground states.
\begin{figure}
\includegraphics[width=9cm, clip]{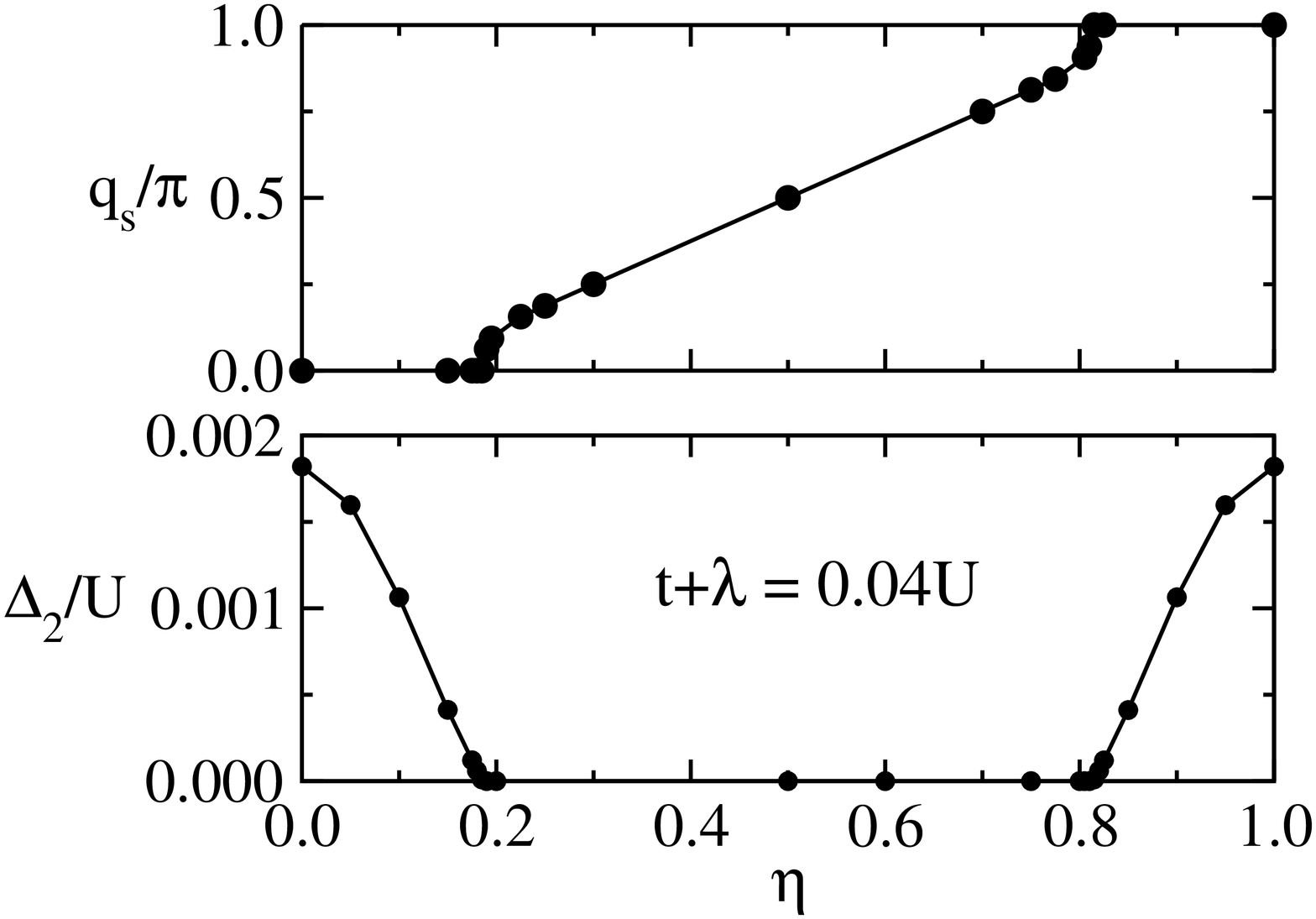}
\caption{The location $q_s$ of the peak of $S^z(q)$ and the longitudinal excitation gap $\Delta_2$ are plotted
as a function of $\eta$ at $t+\lambda=0.04U$ and $U^\prime = 1.2U$.}
\label{FIG4}
\end{figure}
In terms of these quantities defined above, the transition point from the gapped FM phase to the gapless 
chiral phase can be determined accurately by scanning $\eta$. In the upper panel of Fig. \ref{FIG4},
we show the location $q_s$ of the peak of $S^z(q)$ as a function of $\eta$, and obviously a transition from the gapped FM to the gapless
chiral phase occurs at $\eta=0.190(2)$ and the other one from the gapless chiral phase to the gapped AF phase occurs at $\eta=0.810(2)$.
In the lower panel of Fig. \ref{FIG4}, we show $\Delta_2$, which has been extrapolated to the thermodynamic limit,
as a function of $\eta$. In both the gapped FM phase and the gapped AF phase, $\Delta_2$ is finite, 
and zero otherwise in the gapless chiral phase. 
From the results of $q_s$ and $\Delta_2$, we can see again that the features of the gapped AF phase can be deduced 
from those of the gapped FM phase according to the symmetry revealed in Eq. (\ref{EXCHANGE}) with respect to $\eta=0.5$. Moreover,
we note that the critical points determined from $\Delta_2$ are in good agreement with those given by $q_s$,
verifying that $\Delta_2$ and $q_s$ together with $\kappa^\nu_{ij}$ sufficiently characterize the intrinsic features 
for the gapless chiral phase as well as the gapped FM (AF) phase. Furthermore, 
these transitions are qualitatively consistent with our previous analysis based on the effective Hamiltonian (\ref{EHEFF}).

\section{Mott Insulator-superfluid transition}
In this section, we study the MI to SF transition, the magnetic structures as well as the momentum 
distribution in SF phases at the unit filling by DMRG.  

\begin{figure}
\includegraphics[width=9cm,clip]{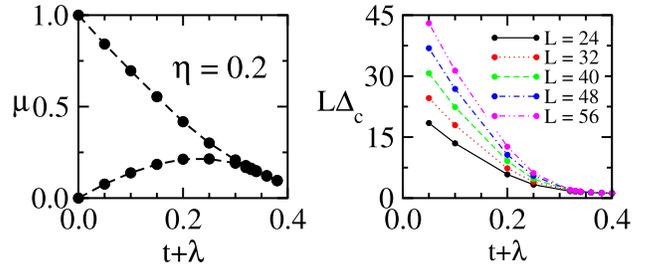}
\caption{(Color online) Left panel: Typical phase diagram for Mott insulator to superfluid transition  
is shown in $(t+\lambda)-\mu$ plane at $\eta=0.2$, $U^\prime=1.2U$. The MI phase is surrounded by the curves and $\mu$-axis. 
The tip of the Mott lobe is estimated to be $t+\lambda \simeq 0.36U$. 
Right panel: Determining the tip of the Mott lobe and criticality by finite-size scaling. $\mu$, $\Delta_c$ and $t+\lambda$
are in unit of $U$.}
\label{FIG5}
\end{figure}
In the left panel of Fig. \ref{FIG5}, we show the phase diagram in $(t+\lambda)-\mu$ plane at $\eta=0.2$, $U^\prime=1.2U$. 
The phase surrounded by the curves and the $\mu$ axis is the MI phase, and outside is the SF phase.
The upper and lower boundaries of the MI lobe are defined\cite{KUHNER1} by the chemical potentials 
$\mu^+=E_0(N+1,L)-E_0(N,L)$ and $\mu^-=E_0(N,L)-E_0(N-1,L)$ in the thermodynamic limit.  
The gap $\Delta_c$ is then defined by $\Delta_c = \mu^+-\mu^-$. The MI phase is characterized by a finite $\Delta_c$.
At $t+\lambda=0$, Hamiltonian (\ref{HSOC}) is decoupled 
into a single-site one and it can be solved exactly, giving $\mu^+=U$ and $\mu^-=0$.  
For other $t+\lambda$, there is no exact solution, and thus we resort to 
DMRG to calculate $\mu^+$ and $\mu^-$ numerically. 
We can see that in the MI phase, $\Delta_c$ is obviously finite and thus the MI phase is incompressible.  
A reentrant behavior for MI-SF transition is observed in Fig. \ref{FIG5}, which is typical in one dimension\cite{KUHNER1}.
We also calculated the phase diagrams at several other $\eta$, and found the curves are basically similar so we 
did not show them. Near the tip of the Mott lobe, the density fluctuation along a constant density line 
is forbidden and only phase fluctuation is allowed. Hence, we expect that such a transition is Berezinskii-Kosterlitz-Thouless(BKT)\cite{BKT}
type. Following the scaling relation\cite{PAI1} 
\begin{equation}   
L\Delta_c\sim f(L/\xi),
\end{equation}
where $\xi$ is the correlation length and $\xi\rightarrow\infty$ in the SF phase, we expect that $L\Delta_c$ separate for different 
$L$ in the MI phase but merge into one curve in the SF phase so that the emerging point gives the critical point. In the right panel,
we show that the transition at the MI tip is indeed BKT type by the finite-size scaling analysis of the charge gap $\Delta_c$
with $L=24,32,40, 48$ and $56$. The tip of the MI lobe is estimated to be $t+\lambda\simeq{0.36U}$. 

In Fig. \ref{FIG1}, we have essentially shown three MI phases which actually differ 
from each other by three different magnetic structures. 
Increasing $t+\lambda$ induces a transition from MI phases to SF phases as seen from Fig. \ref{FIG5}. 
It is natural to wonder whether the magnetic structures survive in the SF phases.  
Below we explore the magnetic structure in the SF phases in terms of correlation functions defined in Eqs. (\ref{SCORR}), (\ref{SQCORR}) 
and (\ref{CHIRAL}) with DMRG calculations. 

\begin{figure}[t]
\includegraphics[width=8.8cm, clip]{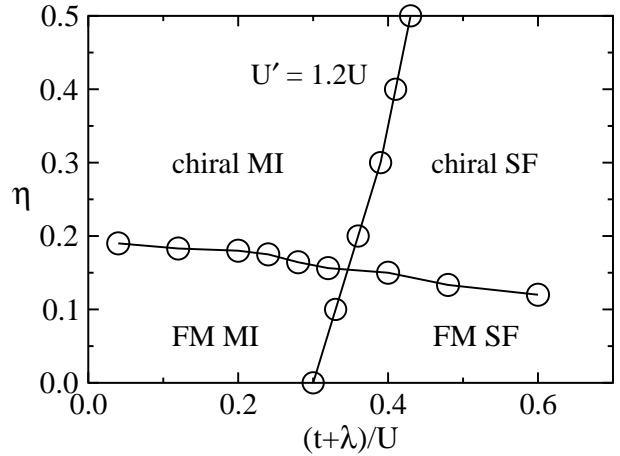}
\caption{A schematic phase diagram is shown in $(t+\lambda)-\eta$ plane for $U^\prime = 1.2U$ at a unit filling.
We show only $0\le\eta\le{0.5}$ due to the symmetry transformation between $\eta$ and $1-\eta$,
see Eq. (\ref{EXCHANGE}). Specifically, by such a transformation,
a point at $(t+\lambda, \eta)$ in the FM MI (SF) phase is transformed into a point at $(t+\lambda, 1-\eta)$ in the AF MI (SF) phase.}
\label{FIG6}
\end{figure}
In Fig. \ref{FIG6}, we show a typical phase diagram in $(t+\lambda)-\eta$ plane at the unit filling and
$U^\prime=1.2U$, which includes four phases: a FM MI phase, a chiral MI phase, a FM SF phase and a chiral SF phase.
The phase boundary between the FM SF and the chiral SF is determined by the peak position $q_s$ of 
the static structure factor defined in Eq. (\ref{SQCORR})
by using the same method as we have discussed in section II.
This phase diagram is plotted only for $0\le\eta\le{0.5}$, since the rest for $0.5\le\eta\le{1}$ can be obtained through 
the transformation (\ref{EXCHANGE}). For a given $\eta$, the phase boundary between the MI and the SF phases is 
just the tip of a Mott lobe as what we show in Fig. \ref{FIG5} so the corresponding transition is of BKT type\cite{BKT}.   
For a given $U^\prime/U$, when $t+\lambda$ increases, quantum fluctuations are enhanced gradually to destroy the FM order
so the area of FM phase shrinks. 
Since the properties of MI phases have been discussed in Section II, we focus on the properties of SF phases below.  

To address the different magnetic properties between the FM SF phase and chiral SF phase,  
\begin{figure}[t]
\includegraphics[width=9cm,clip]{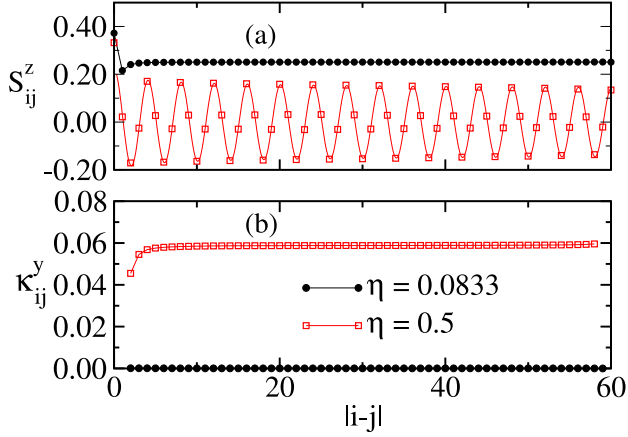}
\caption{(Color online) Typical correlation functions $S^z_{ij}$ and $\kappa_{ij}^y$ are shown as a
function of $|i-j|$ at $t+\lambda = 0.48U$ in the FM SF phase ($\eta = 0.0833$) and the chiral SF phase ($\eta = 0.5$), respectively.
Panel (a) : In the FM SF phase, $S^z_{ij}$ is a finite constant when $|i-j|>4$. In the chiral SF phase, the open squares
are well fitted by the function $S^z_{ij} = 0.1928\times\cos(|i-j|\times\pi/2-0.17)/|i-j|^{0.0696}$, shown as red solid line.
Note that the exponent of $|i-j|$ is very small, indicating a slow algebraic decay in the chiral SF phase.
Panel (b) : In the FM SF phase, $\kappa_{ij}^y$ is exponentially small while in the chiral SF phase it is nearly a finite constant when
$|i-j|>5$. This figure clearly demonstrates the existence of a long-range FM order in the FM SF phase and chiral order
in the chiral SF phase.}
\label{FIG7}
\end{figure}
we show $S^z_{ij}$ and $\kappa^y_{ij}$ in Fig. \ref{FIG7}. 
As we show in panel (a), $S^z_{ij}$ is finite in large $|i-j|$ limit at $(\eta, t+\lambda)=(0.0833, 0.48U)$ belonging to the FM SF phase, 
indicating the emergence of long-range FM order. On the other hand, at $(\eta, t+\lambda)=(0.5, 0.48U)$ for the 
chiral SF phase, $S^z_{ij}$ decays in a power-law
form as a function of $|i-j|$, which can be well fitted by the same function (\ref{SZZFIT}) as that in the chiral MI phase.
Moreover, the relation $q_s=\pi/2$ for the peak position at $\eta=0.5$ remains satisfied in the chiral SF phase, which is strictly protected by 
the symmetric transformation (\ref{EXCHANGE}). 
In panel (b), we show $\kappa^y_{ij}$ as a function of $|i-j|$. As expected, $\kappa^y_{ij}$ is exponentially small in the FM SF phase
but nearly a finite constant in the chiral SF phase, revealing a long-range chiral order. 
\begin{figure}
\includegraphics[width=11cm, clip]{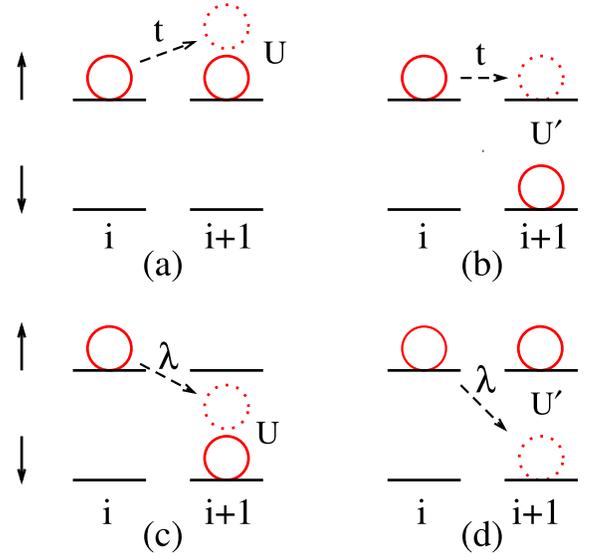}
\caption{(Color online) Schematic picture to explain the magnetic order in the SF phases. Red solid circles 
represent a possible configuration of the ground state, and the arrow indicates a hopping process.  
In (a) and (b), the kinetic energy $t$ dominates,  and FM order (a) is energetically favored. While in 
(c) and (d), the SOC $\lambda$ dominates, and AF order (c) is energetically favored.}
\label{FIG8}
\end{figure}
Since strikingly different from the MI phases, the SF phases at least involve one of the kinetic energy term and the SOC term being 
comparable or even dominant over the interactions. It is interesting to understand how the motion process 
of bosons ensures the magnetic structures. For this purpose, we sketch a simple picture shown in Fig. \ref{FIG8}. 
When $t\gg\lambda$, the SOC might be neglected. A polarized configuration, as we show in panel (a), is energetically 
favored because $U^\prime$ is larger than $U$. In this case, it is effectively a one-component system. 
The particle can only hop within the same component. When $\eta$ increases, 
the probability of hopping between different components increases as well, eventually destroying the FM order.
The strong competition between the kinetic energy and the SOC results in a chiral SF phase. 
As $\eta$ approaches $1$, i.e., $\lambda\gg{t}$, only particle tunneling between different components is allowed, 
and thus configuration in panel (c) is energetically favored, giving rise to a AF SF phase.

\begin{figure}
\includegraphics[width=8.8cm,clip]{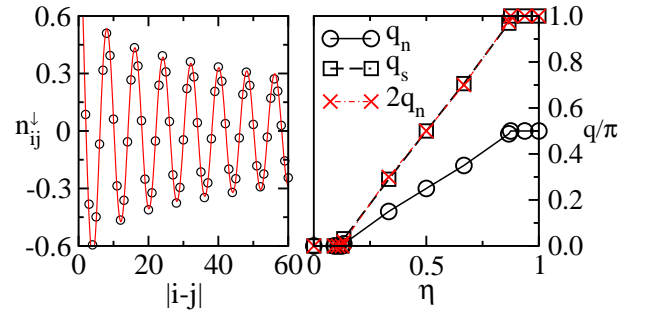}
\caption{(Color online) Left panel: typical one-body density matrix $n^\downarrow_{ij}$ is plotted as a function $|i-j|$ 
at $\eta=0.5,t+\lambda=0.48U$ and $U^\prime=1.2U$. Red solid line is fitting for numerical data, see text. 
Right panel: the peak position $q_n$ of $n^\downarrow(q)$ and 
$q_s$ of $S^z(q)$ are plotted as a function of $\eta$ for $t+\lambda=0.48U$, $U^\prime=1.2U$. Open circle is for $q_n$, open square is for 
$q_s$, and red cross is for $2q_n$.}
\label{FIG9}
\end{figure}
Finally, we elucidate the SF nature of the FM and the chiral SF phases by calculating the one-body density matrix which is defined by 
\begin{equation}
n^\sigma_{ij} = \langle\psi_0| \hat c^{\dagger}_{i \sigma} \hat c_{j \sigma}|\psi_0\rangle,
\end{equation}
and the momentum distribution is given as  
\begin{equation}
n^\sigma(q)=\frac{1}{L}\sum_{ij}e^{iq(i-j)}n^\sigma_{ij},
\end{equation}
where $|\psi_0\rangle$ is the ground state of Hamiltonian (\ref{HSOC}).
In one dimension, there is no true condensation due to strong quantum fluctuation. 
Instead, the SF is characterized by the power-law behavior of the one-body density matrix $n^\sigma_{ij}$,  
\begin{eqnarray}
n^\sigma_{ij}\sim \cos(|i-j|q_n+\delta_n)/|i-j|^{\alpha_n}.
\label{NFIT}
\end{eqnarray}
where $q_n$ corresponds to the peak position of $n^\sigma(q)$.
This formula is applicable for all SF phases. In particular, in conventional SF phases, $q_n=0$, or $\pi/2$, corresponding 
to the FM SF or AF SF. While in the chiral SF phase, $q_n\in(0,\pi/2)$. 
This incommensurability for $n^\sigma(q)$ results from the 
competition between the kinetic energy and the SOC. In the left panel of Fig. (\ref{FIG9}), we show $n^\downarrow_{ij}$ 
as a function of $|i-j|$ at $(\eta, t+\lambda) = (0.5, 0.48U)$, which belongs to the chiral SF phase. 
In this case, fitting the numerical data with Eq. (\ref{NFIT}) gives rise to $n^\downarrow_{ij}= \cos(|i-j|\pi/4-0.116)/|i-j|^{0.3}$.
We noticed that $q_n=\pi/4$, which is half of $q_s$ obtained by fitting $S^z_{ij}$ of Fig. \ref{FIG7} with Eq. (\ref{SZZFIT})
at the same parameters.
Indeed, Eq. (\ref{NFIT}) is directly related to Eq. (\ref{SZZFIT}) and $q_n/q_s=1/2$ always holds, not just at $\eta=0.5$. 
This relation is verified for more $\eta$ by the DMRG calculations in the right panel of Fig. \ref{FIG9},
where $q_n$, $q_s$ and $2q_n$ are plotted as a function of $\eta$. Clearly, the data for $2q_n$ and $q_s$
fall into  one curve, providing strong evidences for our conclusion. This is presumably because  
there are two pairs of operators ($c_{i\tau}$ or $c^\dagger_{i\tau}$) in $S^z_{ij}$ but one in $n^\downarrow_{ij}$. 
Moreover, in the chiral SF phase, $q_n$ and $q_s$ depend almost linearly on $\eta$.
These results may reflect a close relation between magnetic structures and SF pattern.
Finally, we remark that the incommensurability of the momentum distribution with $q_n\in(0,\pi/2)$ 
signals the difference between the chiral SF phase and 
the FM (AF) SF phase. The former has never been reported in TBHM\cite{HU1}, but shows up in 
a wide region in the presence of the SOC,
demonstrating a peculiar feature arising from the competition between the kinetic energy and the SOC.  

\section{CONCLUSIONS}
In conclusion, we studied the evolution of the magnetic structure under a synthetic spin-orbit coupling in 
one-dimensional two-component Bose-Hubbard model with $U^\prime > U$ by using density-matrix renormalization group method. 
When $t+\lambda$ is small and at the unit filling, three magnetic MI phases are found: a gapped FM phase, a gapless chiral phase and 
a gapped AF phase. These magnetic orders persist to SF phases, leading to three different SF phases.  
In particular, in the chiral MI phase and chiral SF phase, the asymptotic behaviors of characteristic 
correlation functions are modulated incommensurately, demonstrating the novel effects 
on the competition between the kinetic energy and the spin-orbit coupling. 
We believe that our findings will inspire further theoretical and experimental investigations on the 
effects of the spin-orbit coupling in lattice bosonic systems.  

\section{ACKNOWLEDGEMENTS}
The computational resources are provided with the high-performance computer-Kohn at physics department, RUC.
P. Zhang is supported by NSFC 91321103, X. Q. Wang is supported by MOST 2012CB921704 and NSFC 11174363.

{\it{Note added}} After we post this paper, three papers on the similar topic appear\cite{PIRAUD1,XU2,PEOTTA1}.

\vfill
\end{document}